\documentclass[twocolumn,showpacs,preprintnumbers,amsmath,amssymb,aps,floatfix]{revtex4}
\usepackage{graphicx}
\usepackage{dcolumn}
\usepackage{bm}

\begin{document}

\title{Ergodicity breaking in strong and network-forming glassy system }
\author{ S. Caponi,$^{1,2}$ M. Zanatta,$^{1}$ A. Fontana,$^{1,2}$
L. E. Bove,$^{3}$ L. Orsingher, $^{1,2}$  F.Natali,$^{4}$ C.Petrillo,$^{5,2}$ F. Sacchetti$^{5,2}$}

\affiliation{
  \\ $^1$Dipartimento di Fisica, Universit\`a di Trento, I-38050 Povo
  Trento, Italy
  \\ $^2$ Research center SOFT-INFM-CNR, Universit\`a di Roma "La Sapienza", I-00185, Roma, Italy
\\$^3$Departement Physique des Milieux Denses CNRS-IMPMC, Universit\`e Paris 6, F-75015 Paris, France
  \\ $^4$ INFM-OGG, and CRS-SOFT, c/o ILL, 6 rue Jules Horwitz, F-38042 Grenoble, Cedex 9, France.
  \\$^5$Dipartimento di Fisica, Universit\`a di Perugia, I-06100 Perugia, Italy.}

\date{\today}

\begin{abstract}

The temperature dependence of the non-ergodicity factor of vitreous GeO$_2$, $f_{q}(T)$, as deduced from elastic
and quasi-elastic neutron scattering experiments, is analyzed. The data are collected in a wide range of
temperatures from the glassy phase, up to the glass transition temperature, and well above into the undercooled
liquid state. Notwithstanding the investigated system is classified as prototype of strong glass, it is found
that the temperature- and the $q$-behavior of $f_{q}(T)$ follow some of the predictions of Mode Coupling Theory.
The experimental data support the hypothesis of the existence of an ergodic to non-ergodic transition occurring
also in network forming glassy systems.

\end{abstract}


\maketitle On cooling a liquid below the melting point of its crystal, its viscosity increases and its molecular
motion is slowed down. Below the glass transition temperature, $T_{g}$, no more molecular rearrangements typical
of a liquid occur: the system, becoming a glass, is completely arrested, at least on the observation time scale.
The structural ($\alpha-$) relaxation process describes the dynamical arrest. Its characteristic time
$\tau_{\alpha}$ exhibits a strong temperature dependence, usually mirrored by that of the static transport
coefficients, such as the shear viscosity, $\tau_{\alpha}\propto\eta$. In the glassy state, only secondary
relaxation processes, $\beta$, remain activated. They are attributed to local reorientational motions.

From the theoretical side, several different microscopic pictures have been proposed to universally describe the
dynamical arrest characterizing the glass transition \cite{congressi, free, MCT, frustr, Debenedetti}. The
existence of a trapping effect that molecules or group of molecules suffer, due to the presence of their
neighbors, was formulated both in the phenomenological approach of the free volume \cite{free} and in the more
formal framework of the Mode Coupling Theory (MCT) \cite{MCT}. The MCT is, up to now, the only approach which
provides a self consistent treatment of the particles dynamics. Thanks to the introduction of a non-linear
microscopic equation of motion, it allows to calculate the time evolution of $\Phi_{q}(t)$, the normalized
density autocorrelation function at momentum transfer $q$. The glass-transition problem is approached
introducing the existence of a dynamic instability at some temperature, $ T_{c}$, above $T_{g}$. The transition
from ergodic (high temperature) to non-ergodic (low temperature) state takes place at this temperature
\cite{MCT}. From the microscopic point of view, $ T_{c}$ corresponds to a critical density from which each
particle is trapped in the cage made by its neighbors. In this picture, only the rattling motion of the
particles blocked inside the cages is permitted: this motion is the microscopic origin of the $\beta$ relaxation
\cite{MCT}.

In recent years, a considerable amount of work has been done to extend and solve the MCT equations for systems
much more complex than simple liquids with spherically symmetric interactions for which the theory was
originally formulated \cite{sciortino,sciortino2,Schilling,Franosch}. Some universal features, largely
independent from the particular system one is analyzing, were foreseen. For instance, in a temperature region
close to $T_{c}$, a cusp-like behavior of the long time limit of the density correlator, $
\Phi_{q}(t)_{\overrightarrow{t\longrightarrow\infty}}f_{q}(T)$ is predicted. In fact, in a liquid $f_{q}(T)$ is
equal to zero, indicating that at long enough times a given particle can leave the transient cage made by its
nearest neighbors. Crossing $T_{c}$, no more molecular rearrangement occurs and $f_{q}(T)$ changes
discontinuously assuming a finite value. In this sense $f_{q}(T)$ is usually considered as an indicator of the
ergodic to non-ergodic transition \cite{MCT}.

The Inelastic Neutron Scattering (INS) and Inelastic X-Ray Scattering (IXS) experiments provide a good mean to
study the temperature evolution of $f_{q}(T)$ in a wide $q$ range. From a experimental point of view, $
\Phi_{q}(t)$ and $f_{q}(T)$ are accessible measuring the intermediate scattering function, $F(q,t)$. In a
supercooled liquid, $ \Phi_{q}(t)$ is expected to go to zero exhibiting a two-step decay related to the $\beta$
and $\alpha$ relaxations respectively. On cooling the sample, the structural rearrangements became slower and
slower, and the decay of $\Phi_{q}(t)$ related to the $\alpha$ relaxation is shifted to longer and longer times.
When the temperature is low enough and the $\alpha$ relaxation is frozen, the characteristic time of its decay
is no more measurable and, for long times, $\Phi_{q}(t)$ assumes a finite value: $f_{q}(T)$. Alternatively, the
plateau value separating the $\beta$ relaxation from the $\alpha$ one, sited at longer times, can be determined
in the frequency space by measuring the dynamic structure factor, $S(q,\omega)$. When the characteristic time of
the $\alpha$ process is shifted to so long time that $1/\tau_{\alpha}$ is inside the frequency interval defined
by the instrumental resolution function, the evolution of the $\alpha$ process can be directly monitored by
measuring the elastic scattering intensity. Under this condition, $f_{q}(T)$ is determined by the ratio of the
elastic to the total scattered intensity, that is $ f_{q}(T)= S_{el}(q,\omega=0)_{T}/S(q)_{T}$.

In the past two decades, the existence of the specific theoretical predictions on the \textit{T} and \textit{q}
behavior of $f_{q}(T)$ motivated a large number of experiments and simulations aimed at testing the MCT in
different systems \cite{MCT2,CaK, frick, bartsch,giulio,daniele}. Recent studies tried also to relate the
microscopic properties of the systems with the temperature dependence of their viscosity, when approaching the
glass transition \cite{buc04,buc92,tullio,lep,bove}. Up to now, the experimental studies performed by INS
\cite{CaK,frick,bartsch} and more recently also by IXS \cite{giulio,daniele} were carried out mainly on glass
formers without directional bounds (Van der Waals or ionic systems) which show high fragility (high increase of
viscosity upon cooling \cite{Angell}). Recently, both theoretical and simulation studies
\cite{sciortino,sciortino2}, and experimental data \cite{comez,corezzi}, have suggested that the MCT can
describe the behavior of $ f_{q}(T)$ also in systems with a spatial organization of the molecules.

The aim of the present work is to experimentally test which MCT predictions, if any, are valid also for strong
and network forming glassy systems. For this purpose, we choose as prototype system the vitreous germania,
v-GeO$_{2}$: it is one of the strongest glass $m_{Geo_{2}}=20$ \cite{Angell2} and it exhibits a rather low glass
transition temperature easy accessible by the experiments ($T_{g} \simeq 800 K$). The samples were obtained by
melt-quench process of the Germanium(IV) oxide crystalline powder purchase by Aldrich (purity greater than
99.998\%). The system has been investigated by neutron scattering experiments.
\begin{figure}[tbp]
\includegraphics[width=9.0cm] {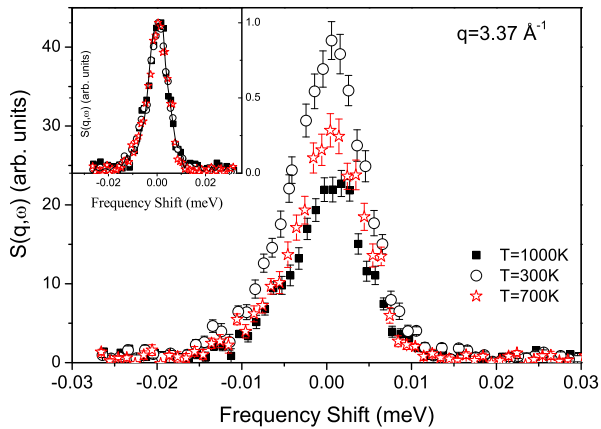}
\caption{Quasi elastic scattering measurement at $q = 3.37 \AA^{-1}$ are reported for selected temperatures, 300
K (circles), 750 K (stars) and 1000 K (squares). In the inset the normalized spectra are reported together with
the instrumental resolution measured by a vanadium scan (full line). The spectra are identical to the
instrumental resolution which, due to the experimental setup, is somewhat asymmetric.} \label{figQE}
\end{figure}

Elastic and quasi-elastic measurements were performed on the thermal backscattering spectrometer IN13 at the
Institut Laue-Langevin (ILL, Grenoble, France). The total elastic intensity, $S_{el}(q,\omega=0,T)$, was
measured as a function of temperature from 20 K to 1100 K. Quasi-elastic spectra were collected at significant
temperatures in the energy range [-0.026 meV to +0.03 meV] with an energy resolution $\Delta \omega = 9 \mu$eV.

The measurements of the static structure factor, $S(q)$, were performed at the diffractometer 7C2 located on the
hot source of the reactor Orph$\acute{e}$e of the Laboratoire Leon Brillouin (LLB, Saclay, France). The incident
wavelength $\lambda =0.729\AA$ was chosen to access a wide $Q$ range between $0.4$ to $15\AA^{-1}$ and to
minimize the inelasticity corrections. The $S(q)$ was investigated in the temperature range from room
temperature to the liquid phase.

The two sets of data were reduced using MonteCarlo simulation to estimate the multiple scattering, the cell
contribution and the transmission coefficients \cite{petrillo}.

\begin{figure}[tbp]
\includegraphics[width=9cm] {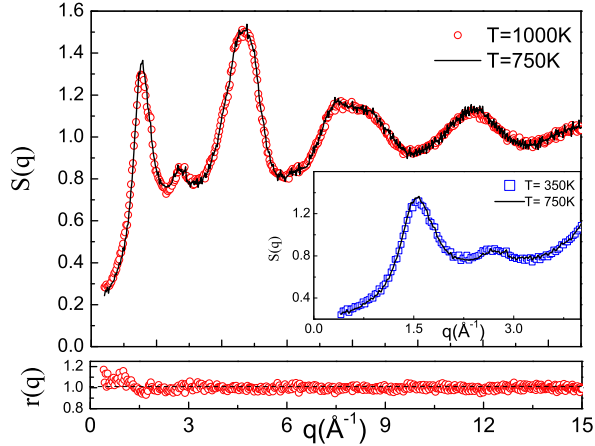}
\caption{Static structure factor $S(q)$ measured at the two relevant temperatures $T$ = 750 K (line, upper
panel) and $T$ = 1000 K (circles, upper panel). The ratio $r(Q) = \frac{S(Q)_{T=1000}}{S(Q)_{T=750}}$ is also
reported (circles, lower panel). The dashed line in the lower panel is an horizontal guide to the eye. The
errors bars are of the same size as the symbols. In the inset the low q part of S(q) measured at T=350K
(squares) and T=750K (line).} \label{figSq}
\end{figure}

In Fig.~\ref{figQE}, where the quasi-elastic spectra are reported for selected temperatures, the decreasing
intensity underlines the temperature effect of the atoms motion. In the inset of Fig.~\ref{figQE} we report the
same spectra, normalized at the maximum intensity. No changes of the quasi-elastic line shape are observed in
the whole explored temperature range, indicating that there are no detectable thermically activated processes,
at least within the width of the experimental resolution. Therefore the $\alpha$ relaxation process is always
inside the elastic peak and the relation $\Delta\omega\tau_{\alpha}(T)>>1$ is always fulfilled. As mentioned
before, under this condition the $f_{q}(T)$ can be analyzed according to the previous description. In
particular, considering that $ f_{q}(T \rightarrow 0) = 1$, $ f_{q}(T)$ is obtained as \cite{bartsch}:
\begin{equation}
\frac{S_{el}(q,\omega=0)_{T}}{S_{el}(q,\omega=0)_{T=0}}
=f_{q}(T)\cdot\frac{S(q)_{T}}{S(q)_{T=0}}
\end{equation}
in this way each data set is normalized by its low temperature measurement, thus cancelling all instrumental
normalization constants. Since our low temperature datum in the elastic scan is $20K$, we linearly extrapolate
the $S(q,\omega=0)_{T=20}$ to $S(q,\omega=0)_{T=0}$. Room temperature is the lowest temperature where the static
structure factor was measured. Since the diffraction pattern does not show any significant change from room
temperature up to 750K (as shown in the inset of Fig.\ref{figSq}), we safely approximate $S(q)_{T = 0}$ to
$S(q)_{T = 350 K}$. Smooth temperature evolution is found in the $S(q)$ for temperatures greater than 750 K. In
Fig.\ref{figSq}, $S(q)$ is presented at two significant temperatures ($T=750$ K and $T=1000$ K) together with
the ratio, $r(q)$, between the two spectra. $r(q)$ is almost constant in the whole range apart at very low $q$,
where slight differences are present.

\begin{figure}[tbp]
\includegraphics[width=7.8cm] {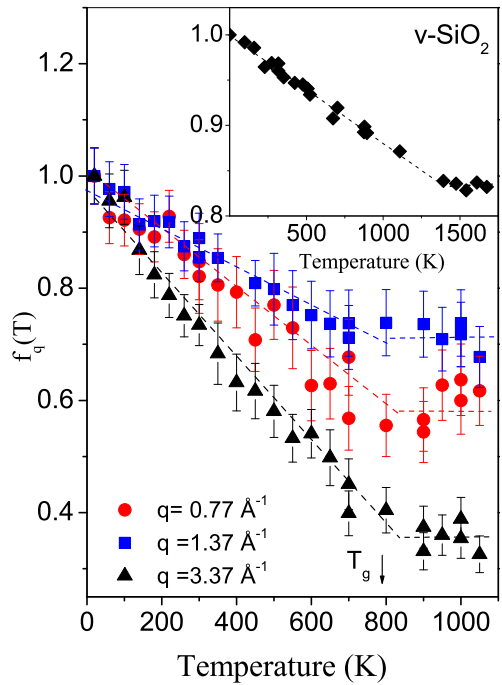}
\caption{Temperature dependence of $f_{q}(T)$ measured in v-GeO$_{2}$ at three selected wave vector transfers.
The dashed lines are reported as guides to the eye. \\ In the inset $f_{q}(T)$ is reported for the v-SiO$_{2}$
sample measured at $q = 2.6 \AA^{-1}$ using the IN6 spectrometer \cite{uli}.} \label{fig2}
\end{figure}

The obtained $f_{q}(T)$ values are reported in Fig.~\ref{fig2} as a function of temperature, for three selected
wave vector transfers, q. The data follow the same trend, without any significant deviation, in the whole $q$
range here investigated. In the low temperature region, the temperature dependence of $f_{q}(T)$ is fairly well
described by a linear behavior. Increasing the temperature a strong slope variation is clearly identified. It
occurs at the same temperature independently from the analyzed $q$ value. In the highest temperature range
$f_{q}(T)$ becomes almost temperature independent at a value which strongly depends on $q$. Interestingly, a
very similar temperature dependence is also found in v-SiO$_{2}$ reported in the inset of Fig. \ref{fig2}. These
measurements were collected using the IN6 spectrometer installed at ILL, with $q \sim 2.6 \AA^{-1}$\cite{uli}.
As previously described, in the MCT, $f_{q}(T)$ is expected to display a cusp behavior in the region close to
the critical temperature, $T_{c}$, which should be located above $T_{g}$. The theory
 foresees that close to $T_{c}$ the following relationship holds:
\begin{equation}
f_{q}(T)= \left \{ \begin{array}{ll} f^{c}_{q} + h_{q}(T-T_{c})^{\beta} &  T<T_{c}
\\  f^{c}_{q} & {\rm T>T_{c}}
\end{array} \right.
\end{equation}
where $\beta = 0.5$ is expected and $ f^{c}_{q}$ and $h_{q}$ are positive quantities depending only on $q$
\cite{MCT}. In order to apply this relation the static structure factor $S(q)$ must depend smoothly on the
temperature: this guarantees that any change in the temperature behavior is not related to a structural
rearrangement, but it has a dynamical origin. This condition is fulfilled by our data as shown in
Fig.\ref{figSq}.
As foreseen by MCT, the presented data suggest the existence of a break in the trend of $f_{q}(T)$ versus
temperature . The strong slope variation in the temperature dependence of $f_{q}(T)$ is located near $T_g$. It
is not straightforward to identify the temperature of the break as $T_c$ defined in Eq. 2 since in strong
glasses it is expected to largely exceed $T_g$ \cite{Sokolov}. However it is evident that a transition occurs
and the system starts to anomalously change its dynamics. It is quite interesting to observe that the same
behaviour is observed also in v-SiO$_{2}$ where $T_c$ is expected to be much higher than the temperature where
the break occurs, i.e. 1500 K. Concerning the temperature dependence of $f_{q}(T)$, in the cases of these two
strong glasses the linear trend of the non-ergodicity factor is maintained up to temperatures close to $T_g$. We
have to recall that Eq.(2) is valid when $T$ is close to $T_{c}$ (probably a few degrees), while the
experimental data span a wide temperature range. Indeed, the data are perfectly compatible with a power law,
with $\beta = 0.5$, at least within some 50 K above $T_g$. This behavior can be obtained taking into account
that the measured $f_{q}(T)$ is the superposition of two contribution: the first related to the relaxational
process described by Eq.(2), the second related to the pure vibrational contribution. The latter, calculated by
the density of vibrational state measured at T=300K \cite{orsingher}, show an almost linear trend in the low
temperature range. However, an accurate analysis in the region close to $T_{c}$ is probably beyond the present
investigation capability in terms of energy resolution, Q range and statistics.

\begin{figure}[tbp]
\includegraphics[width=9cm] {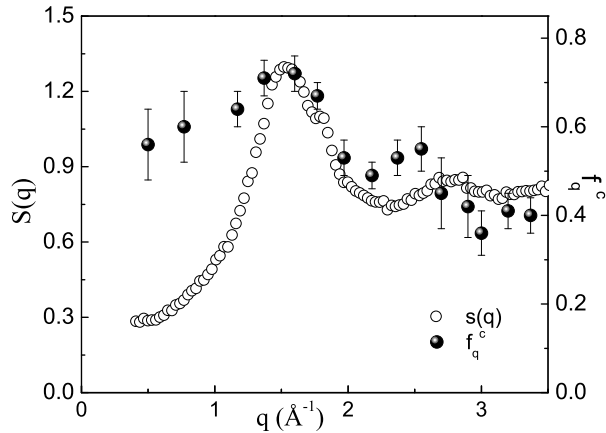}
\caption{Wave-vector dependence for the $f^{c}_{q}$ parameter (dots), compared with the static structure factor,
$S(q)$, of v-GeO$_{2}$ (circles).} \label{fig3}
\end{figure}

A further test to ensure that the constant value reached by $f_{q}(T)$ in the high temperature range is really
$f^{c}_{q}$ of eq.2, is represented by its $q$ dependence. According to MCT calculations for simple liquids,
$f^{c}_{q}$ has to follow in phase the oscillations of the static structure factor, $S(q)$ \cite{MCT}. This
behavior has been experimentally observed in different glass-forming systems \cite{MCT2}. The comparison between
the $q$ behavior of $S(q)$ and $f^{c}_{q}$ obtained by our data is reported in Fig. \ref{fig3}. Again, also in
the present case the ability of $f^{c}_{q}$ to oscillate in phase with the main feature of $S(q)$ is confirmed.

Our results confirm that, also in the analyzed system, the MCT is able to qualitatively describe some universal
aspects of the glass transition. Even if the temperature, at which the ergodicity breaking occurs, is
astonishing close $T_{g}$, its existence is observed in v-GeO$_{2}$ and seems to be visible also in v-SiO$_{2}$.
The existence of the ergodicity breaking in network forming glassy systems has been previously formulated
analyzing the results on B$_{2}$O$_{3}$ where the $f_{q}(T)$ was investigated by different experimental
techniques \cite{brodin}. However, in this latter case, $f_{q}(T)$ does not show a well-defined discontinuity in
its temperature dependence but only a gradual change in a temperature range about 200 K wide. Our findings
suggest that MCT predictions could be extended to a wide class of materials, such as network forming glassy
system. This result is somehow unexpected, in fact these systems are characterized by a local structure
determined by covalent bounds and the significate of caging effect on these materials is far from being trivial.

We acknowledge R. Dal Maschio for the help in the preparation of v-GeO$_{2}$ samples, and U. Buchenau for
helpful discussions and for having provide us the silica data. We thank ILL for providing the neutron beam under
an agreement between ILL and CNR (Italy) and for technical support. The experiment at LLB was supported by the
European Commission through the Access Activities of the Integrated Infrastructure Initiative for Neutron
Scattering and Muon Spectroscopy (NM13), supported by the European Commission under the 6th Framework Programme
through the Key Action: Strengthening the European Research Area, Research Infrastructures, Contract n°:
Rll3-CT-2003-505925.


\begin{thebibliography}{99}

\bibitem{congressi} A. Fontana, P. Verrocchio, G. Viliani (guest editors), 10th International Workshop on Disordered Systems, Andalo, 2006, Phil. Mag. {\bf B 87} (2007).
\bibitem{free}M.H. Cohen, G.S. Grest, Phys. Rev. B {\bf 20}, 1077 (1979).
\bibitem{MCT} W. G\"{o}tze, Liquids, Freezing and the glass transition, edited by J. P. Hansen, D. Lesvesque, and J. Zinn-Justin, North-Holland, Amsterdam (1991); W. G\"{o}tze, and M. R. Mayr, Phys. Rev. E{\bf 61} 587 (2000); W. G\"{o}tze and L. Sj\"{o}gren, Rep. Prog. Phys. {\bf 55}, 241 (1992).
\bibitem{frustr}D.Kivelson et al., Physica A {\bf219}, 27 (1995); P.Viot, G.Tarjus, D.Kivelson, J. Chem. Phys. {\bf 112}, 10368 (2000).
\bibitem{Debenedetti} P.G. Debenedetti and F.H. Stillinger, Nature (London) {\bf 410}, 259 (2001).
\bibitem{sciortino} F. Sciortino, W. Kob, Phys. Rev. Lett. {\bf 86}, 648 (2001).
\bibitem{sciortino2} L. Fabbian et al., Phys. Rev. E {\bf 60}, 5768 (1999).
\bibitem{Schilling} R.Schilling, T.Scheidsteger, Phys.Rev.E {\bf 56}, 2932 (1997).
\bibitem{Franosch}T. Franosch et al., Phys. Rev. E {\bf 56}, 5659 (1997).
\bibitem{MCT2}W. G\"{o}etze J. Phys. Cond. Matt. {\bf 11}, A1 (1999).
\bibitem{CaK} E. Kartini et al., Phys. Rev. B {\bf 54}, 6292 (1996).
\bibitem{frick} B. Frick, B. Farago, D. Richter, Phys. Rev. Lett. {\bf 64}, 2921 (1990).
\bibitem{bartsch} E. Bartsch et al., Phys. Rev. E {\bf 52}, 738 (1995).
\bibitem{daniele} D. Fioretto et al., Phys. Rev. B {\bf 65}, 224205 (2002).
\bibitem{giulio}G. Monaco, C. Masciovecchio, G. Ruocco, and F. Sette, Phys. Rev. Lett. {\bf 80}, 2161 (1998).
\bibitem{tullio}T. Scopigno, G. Ruocco, F. Sette, and G. Monaco, Science {\bf 302}, 849 (2003).
\bibitem{buc04}U. Buchenau and A. Wischnewski, Phys. Rev. B {\bf 70}, 092201 (2004).
\bibitem{buc92}U. Buchenau, R. Zorn, Europhys. Lett. {\bf 18}, 523 (1992).
\bibitem{lep}L. Larini, A. Ottochian, C. De Michele, D. Leporini, Nature Physics {\bf 4} 42 (2008).
\bibitem{bove} L.E. Bove, C. Petrillo, A. Fontana, A. P. Sokolov, J. Chem. Phys. \textbf{128} 184502 (2008).
\bibitem{Angell} For a review, see M.C. Ediger, C.A. Angell, and S.R. Nagel, J. Phys. Chem. {\bf 100}, 13 200 (1996).
\bibitem{comez} L. Comez et al., Phys. Rev. Lett. {\bf 94}, 155702 (2005).
\bibitem{corezzi} S. Corezzi et al., Phys. Rev. Lett. {\bf 96}, 255702 (2006).
\bibitem{petrillo}C. Petrillo, F. Sacchetti, Acta Cryst. {\bf A46}, 440 (1990); C. Petrillo, F. Sacchetti, Acta Cryst. {\bf A48}, 508 (1992).
\bibitem{Angell2} R. B\"{o}hmer, K.L.Ngai, C.A. Angell, and D.J.Plazek, J. Chem. Phys. \textbf{99}, 4201 (1993).
\bibitem{uli} These data are kindly supplied by U.Buchenau. \\ A. Wischnewski thesis umpublised.
\bibitem{Sokolov} V.N. Novikov, E. Rossler, V.K. Malinovsky, N.V. Surovtsev, Europhys. Lett. 35,  289, (1996).
\bibitem{orsingher} L. Orsingher et al. to be published.
\bibitem{brodin} A. Brodin, et al. Phys. Rev. B {\bf 53}, 11511 (1996); D. Sidebottom, R. Bergman, L. B$\ddot{o}$rjesson, and L. M. Torell, Phys. Rev. Lett. {\bf 71}, 2260 (1993).

\end{thebibliography}
\end{document}